7,849 words in the main text

232 words in the abstract

62 references

1 table and 8 figures in the main text

# TransCAVE–E: A distributed virtual reality testbed for adaptive external human-machine interfaces

Yun Ye[1,2], Zexuan Li[2], Haoyang Liang[3,4,✉], Boya Sun[4], Jian Sun[4], Haotian Shi[4,✉]

[1] Centre for Global Infrastructure Resilience, The Bartlett School of Sustainable Construction, University College London, London WC1E 7HB, UK

[2] Faculty of Maritime and Transportation, Ningbo University, Ningbo 315211, China

[3] College of Transport and Communications, Shanghai Maritime University, Shanghai 201306, China

[4] College of Transportation, Key Laboratory of Road and Traffic Engineering, Ministry of Education, Tongji University, Shanghai 201804, China

✉ Corresponding author.

E-mail: hyliang@shmtu.edu.cn; shihaotian95@tongji.edu.cn

**Acknowledgments**

This research was supported by grants from the National Natural Science Foundation of China (Project No. 52302378, 52125208, 72501150), Zhejiang Provincial Natural Science Foundation of China (Grant No. LQN25E080011), Ningbo Natural Science Foundation (Grant No. 2024J440). The experiment conducted in this study was facilitated by TransCAVE Lab (https://github.com/TSILAB/TransCAVE-E) at Tongji University.

## ABSTRACT

External human–machine interfaces (eHMIs) are evolving from predefined displays toward adaptive communication strategies that respond to changing traffic and road-user states. This transition requires experimental infrastructure that can jointly support human-in-the-loop (HIL) interaction, software-in-the-loop (SIL) algorithm execution, reusable experiment orchestration, and synchronized multimodal human-factors evaluation. This paper presents *TransCAVE–E*, a distributed virtual-reality testbed designed for the development and validation of adaptive and intelligent eHMIs. The platform integrates three tightly coupled modules: a Scenario Design Center for configurable traffic environments, task flows, and reusable experimental conditions; an eHMI Algorithm Module that enables bidirectional real-time coupling between the simulation and independent external algorithms; and a Data Management System that synchronizes behavioral trajectories, eye-tracking, physiological, system-log, and subjective data. A distributed multi-agent architecture further supports synchronous interaction among pedestrians, human drivers, automated vehicles, and other traffic entities. Two representative use cases demonstrate the platform's applicability. In the AV–pedestrian experiment, an intent-recognition-based eHMI improved decision efficiency by 12.8% and 13.0% in yielding and non-yielding scenarios, respectively, reduced gaze distraction by 17.1% in the yielding scenario, and reduced unnecessary eHMI prompts by 40% in the non-yielding scenario while maintaining overall interaction safety. A second HV–AV study further demonstrated real-time SIL validation of a game-theoretic information-disclosure strategy under active driver interaction. These results establish *TransCAVE–E* as an extensible and reproducible infrastructure for closed-loop evaluation and iterative refinement of intelligent human–vehicle communication strategies.

## 1. Introduction

With the rapid advancement of highly automated vehicles (AVs), vehicles are increasingly evolving from isolated execution units into social driving agents that must interact and make interdependent decisions with human-driven vehicles (HVs) and vulnerable road users (VRUs) (Ahmed et al., 2022; Li et al., 2023). External human–machine interfaces (eHMIs) have consequently emerged as an important means of supporting communication between AVs and surrounding road users (Mourtzis et al., 2023; Zhang et al., 2022). Conventional eHMIs primarily communicate predefined vehicle states or intentions through visual or auditory information. However, their effectiveness becomes increasingly context-dependent in mixed and complex traffic environments involving multiple vehicles and road users (Mandujano-Granillo et al., 2024; Tang et al., 2025; Yu et al., 2022). Ambiguity regarding the intended recipient, inconsistency between communicated information and vehicle behavior, and variations in interaction context may lead to misunderstanding or delayed responses (Kato et al., 2026; Cai et al., 2025). These limitations are motivating a transition toward adaptive and context-aware eHMIs capable of modifying communication according to evolving traffic situations and road-user states.

This transition fundamentally changes the object of eHMI evaluation from a predefined interface to a closed-loop interaction strategy. Conventional eHMI evaluation can typically be conducted by presenting fixed interface conditions and comparing road-user responses. Intelligent eHMIs, in contrast, may need to acquire the current interaction state, interpret the behavior or inferred intention of surrounding road users, determine whether and how communication should occur, and dynamically adapt communication as the interaction evolves (Meo et al., 2025; Avetisyan et al., 2025; Tabone et al., 2023). Human responses to these communication strategies can subsequently alter the interaction state and affect later system decisions. Evaluation therefore needs to consider not only the appearance or comprehensibility of an interface, but also the coupling among traffic context, algorithmic decision-making, vehicle behavior, communication strategy, and human response. This creates a growing need for experimental infrastructure capable of supporting reproducible closed-loop interaction, continuous validation, and iterative strategy refinement (Yamin et al., 2023; Guo et al., 2022).

Simulation and virtual reality (VR) provide a safe, controllable, and repeatable basis for such evaluation, particularly when complex or safety-critical human–vehicle interactions need to be reproduced experimentally. Nevertheless, intelligent eHMI testing introduces several requirements that are not easily addressed within conventional study-specific experimental workflows. First, experiment configuration needs to extend beyond static scene construction toward task-oriented orchestration of interaction processes. Researchers must be able to specify not only roads, vehicles, and environmental conditions, but also interaction events, participant tasks, conflict points, temporal sequences, and combinations of experimental conditions in a reusable form. Second, eHMI evaluation needs to extend beyond predefined communication conditions toward closed-loop HIL–SIL execution. Real-time traffic and participant states should be accessible to independent and replaceable communication or decision algorithms, whose outputs can be returned to the simulation and presented to human participants without interrupting the interaction process. This requirement becomes more demanding in multi-user settings, where heterogeneous human participants and traffic entities must remain synchronized within the same environment. Third, multimodal human-factors evidence needs to be integrated within the experimental workflow. Although trajectories, eye movements, physiological responses, and subjective assessments are increasingly collected in human–vehicle interaction studies, these data are often acquired through heterogeneous devices and software pipelines. Their synchronized association with experimental conditions, interaction events, eHMI states, algorithm outputs, participants, and trials are essential for reproducible evaluation and iterative optimization.

To address these requirements, this paper presents *TransCAVE–E*, a distributed simulation-based test platform for the design and evaluation of adaptive and intelligent eHMI strategies, as illustrated in Fig. 1. Rather than introducing another general-purpose driving simulator, *TransCAVE–E* provides an experiment-oriented integration layer that combines HIL and SIL evaluation within a unified workflow. The Scenario Design Center (SDC) supports highly customizable traffic environments together with task-oriented interaction-flow design and reusable experimental configurations. The eHMI Algorithm Module (EAM) provides standardized bidirectional interfaces through which real-time scene and participant states are transmitted to independent external algorithms and resulting eHMI commands are returned to the virtual environment. A distributed multi-agent architecture enables pedestrians, human drivers, AVs, and other traffic entities to interact synchronously within a shared simulation environment. The Data Management System (DMS) further integrates behavioral trajectories, eye-tracking measures, physiological signals, experimental logs, and subjective assessments within a common data structure. Together, these components establish a closed-loop workflow linking experiment configuration, intelligent strategy execution, human interaction, multimodal data acquisition, and iterative evaluation.

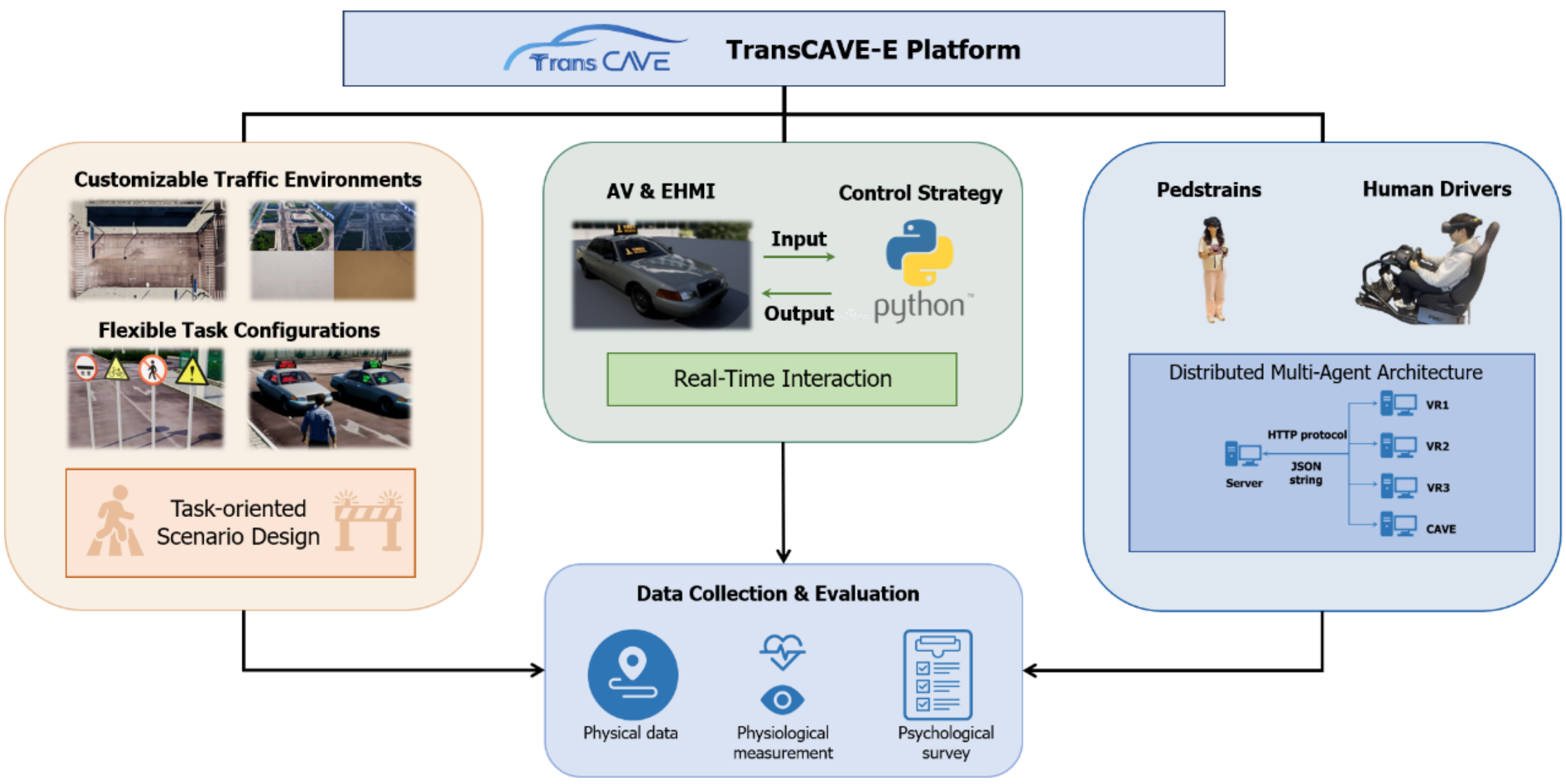

**Fig. 1** *TransCAVE–E* platform overview

The remainder of this paper is organized as follows. Section 2 reviews the evolution of intelligent eHMI and existing human-centered and automated-driving simulation environments, and identifies the need for an integrated HIL–SIL testbed. Section 3 presents the architecture and workflow of *TransCAVE–E*, Section 4 describes its key features, Section 5 demonstrates the two representative use cases, and Section 6 concludes the paper.

## 2. Related work

This section reviews recent studies on eHMI, HIL experimentation, and simulation-based and distributed platforms, with particular attention to the experimental capabilities required for adaptive and algorithm-driven eHMI evaluation. Based on this review, the platform-level research gaps addressed by TransCAVE–E are summarized.

### 2.1 eHMI

Early eHMI research primarily treated external communication as a display-design problem. Studies examined how communication effectiveness varied with factors such as color, position, message content, presentation modality, activation distance, and vehicle-yielding behavior (Bazilinskyy et al., 2021). Within this paradigm, an eHMI was typically defined as an experimental condition whose appearance and triggering logic were specified before an interaction took place. Such designs have generated substantial knowledge regarding interface comprehensibility and pedestrian responses, but they largely assume that the appropriate communication strategy can be determined in advance. This assumption becomes increasingly restrictive in complex traffic environments. The effectiveness of an eHMI can depend on the behavior of surrounding vehicles, the identity and actions of other road users, and consistency between the communicated message and the vehicle's actual motion. In multi-vehicle or multi-road-user situations, external communication may even generate misleading interpretations rather than uniformly improving interaction (Ye et al., 2026). Dynamic eHMIs may improve visibility, comprehensibility, or perceived safety, yet inconsistency between external messages and vehicle kinematics can introduce additional uncertainty and adversely affect pedestrian decisions (Kang et al., 2025; Lau et al., 2022). These findings indicate that eHMI effectiveness is not solely a property of the visual or auditory interface itself; it emerges from the relationship among communication content, interaction context, vehicle behavior, and human response.

Accordingly, recent research has begun to conceptualize eHMI as an adaptive communication system rather than a static display. Tran et al. (2025)

distinguished adaptive external communication from communication that merely reacts to predefined triggering conditions and proposed a conceptual architecture linking information input, decision processing, and communication output. Under this perspective, an eHMI should be able to modify its communication according to changing road-user states and traffic contexts rather than repeatedly execute a fixed mapping between a predetermined event and a predetermined message. This represents a fundamental change in the object being evaluated: the research question shifts from which interface is better? toward what should the system communicate, to whom, and at what moment under an evolving interaction state? Other recent studies have moved further toward algorithm-driven eHMI design. Colley et al. (2025), for example, incorporated human behavioral performance and subjective evaluations into a multi-objective Bayesian optimization process for visual and auditory eHMI parameters, illustrating how human feedback can contribute directly to iterative interface optimization. Xia et al. (2025) further combined large language models with three-dimensional rendering to automate the generation of executable eHMI actions according to intended communication objectives and incorporated human evaluations and vision–language models into the assessment process. Collectively, these developments indicate a transition from fixed communication, through state-triggered communication, toward adaptive and algorithmically generated interaction strategies.

This transition also changes the requirements placed on experimental infrastructure. Evaluation of a predefined eHMI can be conducted by presenting several fixed conditions and comparing participants' responses. Evaluation of an adaptive eHMI, however, requires a continuous interaction loop in which the current traffic and human states are acquired, an interaction or communication algorithm processes these states, the resulting eHMI strategy is returned to the experimental environment, and the road user responds to the updated communication. The resulting behavioral response may then further alter the interaction state. An experimental testbed for intelligent eHMI therefore needs to support not only interface presentation but also configurable interaction processes, real-time algorithm execution, human-in-the-loop interaction, and synchronized evaluation of the resulting human responses.

### 2.2 Human-in-the-loop experiments

Human-in-the-loop experimentation has been central to eHMI research because the effectiveness of external communication ultimately depends on how human road users perceive, interpret, and act upon the information provided. Existing studies have employed a broad range of experimental approaches, including pre-rendered stimuli, desktop-based virtual environments, immersive VR, CAVE systems and driving simulators.

Online and pre-rendered experimental paradigms provide a highly controllable and scalable means of exploring large eHMI design spaces. Bazilinskyy et al. (2021), for example, systematically varied multiple eHMI and traffic-related factors across hundreds of pre-rendered conditions. Such methods are well suited to factorial evaluation of display characteristics, but vehicle trajectories, environmental conditions, and eHMI states are generally generated before the experiment and therefore cannot respond dynamically to the participant's behavior. Similarly, customized virtual environments allow researchers to manipulate eHMI concepts and contextual factors under controlled conditions, but their interaction processes are typically produced or controlled specifically for the study being conducted. Immersive VR has substantially extended this paradigm by allowing participants to experience traffic interactions from a first-person perspective while researchers retain experimental control. Existing VR studies have manipulated vehicle yielding behavior, vehicle size, eHMI design, activation timing, and traffic context (Fratini E et al., 2023; De Clercq et al., 2019). CAVE-based environments further provide larger physical movement spaces and allow participants to exhibit more natural crossing behavior. Lee et al. (2022), for example, used a CAVE-based pedestrian simulator with head-position-responsive rendering, while Kaleefathullah et al. (2022) investigated pedestrian responses under yielding, non-yielding, and misleading eHMI conditions. More complex virtual environments have also incorporated multiple vehicles and multiple road users (Tran et al., 2024). These developments have progressively increased experimental immersion and scenario complexity. However, increasing visual or interaction complexity does not necessarily result in a reusable experimental infrastructure. In many existing systems, vehicle behavior, eHMI triggering logic, experimental events, and interaction sequences remain implemented through scripts written specifically for an individual study. Consequently, changing the research question may require substantial reconstruction of the experimental logic. The distinction is important: a system may support a complex experimental scenario without providing a general-purpose mechanism for configuring and reusing experimental task flows.

Human-in-the-loop experimentation becomes more demanding when participants actively control traffic agents. Driving-simulator studies of HV–AV communication require continuous driver inputs to be synchronized with AV behavior and eHMI states. Rettenmaier et al. (2019; 2020) investigated eHMI communication in road-bottleneck interactions using static and dynamic driving simulators, demonstrating mature support for driver inputs, vehicle dynamics, and controlled communication conditions. Avetisyan et al. (2025) further combined a VR driving simulator with eye tracking and subjective measures of situation awareness, trust, acceptance, and workload, while other studies have examined vehicle-mounted and infrastructure-based eHMIs at intersections (Lingam et al., 2024) and external communication for automated trucks (Gwak et al., 2022). Networked and multi-participant experiments further extend human-in-the-loop research toward interactive multi-agent settings. Hübner et al. (2022), for example, connected driving-simulation and pedestrian-VR components so that a human driver and a pedestrian could participate synchronously in the same interaction. Such work demonstrates the feasibility and value of distributed human interaction. Nevertheless, cross-system synchronization and entity mapping are commonly developed for the specific experimental configuration being studied. Extending the same implementation to different combinations of human roles, traffic entities, or experimental tasks may therefore require additional system-level customization.

Human-factors measurement has evolved in parallel. Beyond trajectories and task-performance measures, eHMI studies increasingly incorporate eye tracking, subjective questionnaires, video recordings, and physiological or workload-related measures (He et al., 2021; Velasco et al., 2019; Deb et al., 2018). These signals provide complementary evidence regarding attention allocation, perceived safety, trust, comprehension, and decision making. The central limitation is therefore not that existing studies fail to collect human-factors data. Rather, scenario configuration, experimental execution, eHMI control, participant sensing, questionnaires, and subsequent analysis are often distributed across separate software and hardware pipelines. The resulting experimental environment is effective for answering a particular research question, but the complete workflow is not necessarily organized as a reusable platform in which scenario conditions, eHMI states, behavioral responses, eye movements, physiological signals, and subjective evaluations are systematically associated with the same experimental trials.

Overall, human-centered eHMI evaluation has progressed from passive stimulus presentation toward immersive, active, and increasingly multi-agent interaction. However, much of the existing infrastructure remains study-centric rather than platform-centric: sophisticated experimental functions are available, but they are commonly assembled around individual research protocols and predefined communication strategies.

### 2.3 Simulation-based and distributed platforms

A second research stream has focused on the development of general simulation platforms for automated driving. High-fidelity simulators such as CARLA (Dosovitskiy et al., 2017) and LGSVL (Rong et al., 2020) provide configurable road environments, vehicle dynamics, traffic agents, sensor interfaces, and access to simulation states, forming an important technical foundation for the development and evaluation of automated-driving systems. Other platforms have placed greater emphasis on multi-agent behavior and large-scale simulation. SMARTS (Zhou et al., 2021), Nocturne (Vinitsky et al., 2022) and MOSS (Zhang et al., 2024), for example, support research on multi-agent interactions, scalable traffic simulation, or learning-based driving strategies. Sim4CV (Müller et al., 2018), SUMMIT (Cai et al., 2020), MACAD (Palanisamy et al., 2020), VRMMO (Shi et al., 2023), and OpenCAMS (Ahmad et al., 2025) further extend the available capabilities for visual simulation, connected and automated mobility, multi-agent control, and co-simulation.

Scenario-generation frameworks have also improved the systematic specification of automated-driving test conditions. Platforms such as MetaDrive (Li et al., 2022) support the composition of diverse driving scenarios, while scenario-description approaches such as Scenic (Fremont et al., 2019) and related specification languages (Zhang et al., 2020) provide mechanisms for defining traffic situations and generating test scenarios. Distributed simulation architectures further make it possible to connect multiple traffic agents or simulation components within a shared environment, which is particularly valuable for interactive and multi-user experiments (Kalantari et al., 2023). These platforms provide many of the technical capabilities required by intelligent eHMI research, including high-fidelity rendering, vehicle dynamics, traffic-state acquisition, multi-agent simulation, network communication, and software interfaces for external algorithms. Nevertheless, their principal design objectives generally concern automated-driving algorithm development or traffic-system simulation, rather than the orchestration of controlled human-subject experiments. A traffic simulator may allow researchers to construct a road, spawn vehicles, modify weather, and execute an external control algorithm, but these functions do not by themselves provide an experiment-oriented workflow for specifying a participant's sequence of tasks, organizing experimental conditions, synchronizing heterogeneous human participants, or associating multimodal human-factors measurements with individual interaction events and trials. The key limitation is therefore not a lack of sophisticated simulation engines. Existing platforms already provide mature underlying capabilities for rendering, physics, traffic-agent control, and algorithm testing. What remains less developed is an experimental layer specifically designed to convert these capabilities into a reusable intelligent-eHMI testing workflow. Such a layer must coordinate the virtual traffic environment with human participants, interaction tasks, external communication algorithms, and human-factors measurements rather than treating these elements as independent components.

This distinction is particularly important when human-in-the-loop (HIL) and software-in-the-loop (SIL) evaluation need to occur simultaneously. Conventional automated-driving SIL testing typically evaluates how an algorithm controls or predicts the behavior of simulated vehicles. Intelligent eHMI testing additionally requires the outputs of an external algorithm to affect the information perceived by a human participant, whose subsequent behavioral response changes the evolving interaction. The test platform therefore needs to maintain a bidirectional relationship between simulation states and communication algorithms while preserving the controlled and reproducible characteristics required for human-subject experimentation.

### 2.4 Research gap

The preceding literature reveals complementary strengths but also a clear gap between existing human-factors experimentation and automated-driving simulation. Human-centered eHMI studies provide controlled and increasingly immersive methods for measuring how road users respond to vehicle communication, whereas automated-driving platforms provide sophisticated simulation, state access, multi-agent coordination, and algorithm execution. However, adaptive and context-aware eHMI increasingly requires these capabilities to operate within the same experimental loop.

Table 1 summarizes representative eHMI experimental environments according to the platform-level capabilities identified above, including task-oriented scenario and workflow configuration, external-algorithm software-in-the-loop integration, distributed or multi-user human-in-the-loop interaction, multimodal human-factors acquisition, and integrated data management. The comparison shows that individual capabilities have already been demonstrated across different studies, but they are typically implemented separately or for specific experimental tasks rather than integrated within a reusable end-to-end testing framework.

Three platform-level requirements arise from this transition. First, intelligent eHMI evaluation requires a shift from predefined communication conditions to closed-loop strategy execution. As discussed in Section 2.1, adaptive and algorithm-driven eHMIs require real-time traffic and participant states to be transmitted to an independent and replaceable algorithm, which determines whether, when, and how communication should occur. The resulting eHMI or vehicle-control commands must then be returned to the simulation environment and presented to the participant in real time. Such external-algorithm software-in-the-loop capability is necessary when the object being evaluated is an adaptive communication strategy rather than a predefined display condition. Second, HIL experimentation requires a shift from study-specific scenario construction toward reusable experiment orchestration. As shown in Section 2.2, existing experimental environments can support complex traffic scenarios, immersive interaction, and even multiple human participants, but interaction events, participant tasks, temporal sequences, conflict points, experimental conditions, and cross-system synchronization are commonly configured specifically for individual studies. A reusable platform should therefore support task-oriented scenario and workflow configuration as well as distributed interaction among heterogeneous human participants within a synchronized shared environment. In this sense, the platform should represent not only where an experiment occurs, but also how the interaction unfolds and how different human-controlled entities participate in it. Third, human-factor evaluation must shift from separate measurements to integrated experimental evidence. As discussed in Section 2.3, existing simulation-based and distributed platforms already provide mature support for traffic-state acquisition, multi-agent simulation, network communication, and external algorithm execution. However, these capabilities are primarily designed for vehicle-algorithm testing or traffic-system simulation, rather than for organizing heterogeneous human-factors measurements within controlled eHMI experiments. Existing eHMI studies already collect trajectory, gaze, physiological, behavioral, and subjective evaluations, but these data are commonly recorded and processed through separate systems. An experiment-oriented platform should therefore synchronize and associate these measurements with the corresponding scenario, interaction event, eHMI state, algorithm output, participant, and trial within a common data structure.

Distributed human interaction provides an additional enabling layer across these requirements. Complex future traffic environments may involve pedestrians, human drivers, AVs, and other agents acting simultaneously. A reusable platform should therefore support heterogeneous human participants within a synchronized shared environment rather than assuming a single participant role. Taken together, these requirements point to a missing experimental layer between general-purpose simulation and study-specific human-factors experimentation. What is needed is not simply another VR environment for comparing predefined eHMI designs, but an integrated infrastructure that combines task-oriented experiment configuration, distributed human-in-the-loop interaction, external-algorithm software-in-the-loop execution, and synchronized multimodal human-factors data management. *TransCAVE–E* is developed to address this gap by integrating these capabilities into a unified workflow for the configuration, deployment, evaluation, and iterative refinement of adaptive and algorithm-driven eHMI strategies.

**Table 1**. Comparison of Experimental Frameworks and Platform Functions in Representative eHMI Studies

| Study / Platform | Task-oriented scenario & workflow configuration | External-algorithm SIL | Distributed / multi-user HIL | Multimodal human-factor acquisition | Integrated data management |
|---|---|---|---|---|---|
| Bazilinskyy et al. (2021) | △ | × | × | × | × |
| Fratini et al. (2023) | △ | × | × | △ | × |
| De Clercq et al. (2019) | △ | × | × | △ | × |
| Lee et al. (2022) | △ | × | × | △ | × |
| Kaleefathullah et al. (2022) | △ | × | × | △ | × |
| Tran et al. (2024) | △ | × | × | △ | × |
| Rettenmaier et al. (2019) | △ | × | × | × | × |
| Rettenmaier et al (2020) | △ | × | × | △ | × |
| Hübner et al. (2022) | △ | × | △ | △ | × |
| Avetisyan et al. (2025) | △ | × | × | △ | × |
| Lingam et al (2024) | △ | × | × | △ | × |
| Gwak et al. (2022) | △ | × | × | △ | × |
| TransCAVE–E | √ | √ | √ | √ | √ |

Note: ✓ indicates explicit and reusable platform-level support; △ indicates a limited or study-specific implementation of the corresponding capability; × indicates that the capability was not used or reported. External-algorithm SIL specifically refers to a closed-loop process in which real-time interaction states are transmitted to an independent external algorithm and the resulting eHMI or vehicle-control commands are returned to the simulation environment. Integrated data management refers to centralized association and management of heterogeneous experimental data across participants, conditions, tasks, timestamps, and data modalities.

## 3. TransCAVE–E workflow

### 3.1 Overview

As illustrated in Fig. 2, *TransCAVE–E* establishes an experimental workflow for adaptive and context-aware eHMI evaluation through three tightly coupled core modules: Scenario Design Center (SDC), EHMI Algorithm Module (EAM) and Data Management System (DMS). Together, these modules form a closed–loop pipeline that links scenario and task construction, real–time eHMI strategy execution, and multimodal human–factors data collection, enabling both real–time feedback and reproducible experimentation.

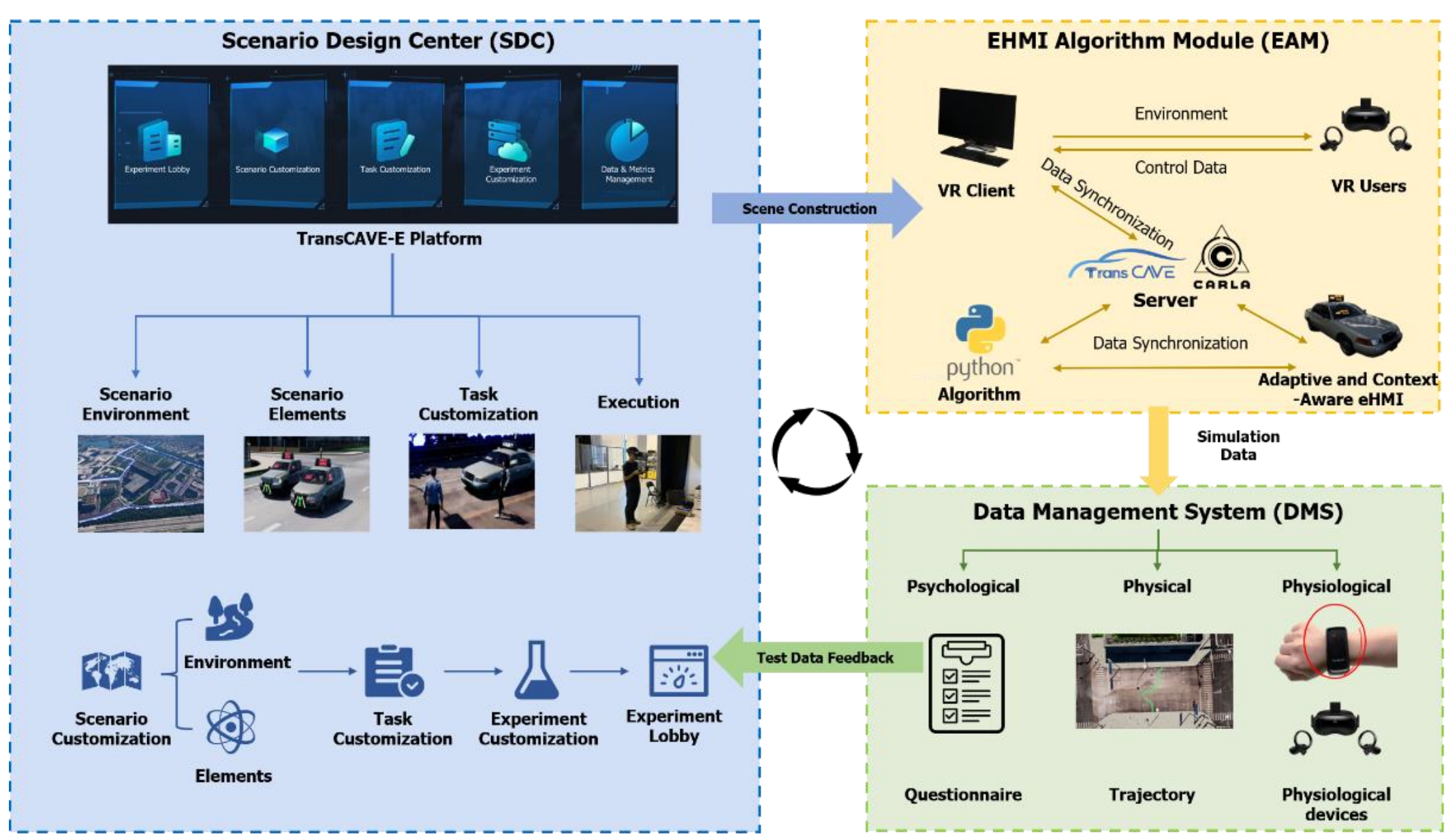

**Fig. 2** Architecture and closed–loop workflow of *TransCAVE–E*.

SDC serves as the entry module of *TransCAVE–E*. By integrating scenario environments and element–model resources, SDC enables researchers to construct high–fidelity virtual traffic environments. On this basis, it supports task–flow customization oriented to experimental objectives and provides unified management of task combinations under different experimental conditions; experiment execution is also initiated from this module through configuration loading and deployment. Based on the scenario and task configurations produced by the SDC, the EAM operates as a SIL component: the Server streams scene and traffic states to the EAM in real time, the EAM continuously generates dynamic eHMI prompt commands and transmits its outputs back to the Server, and the prompts are rendered and presented in real time via the VR/CAVE Clients. During this process, interaction logs and human–factors data are automatically aggregated into the DMS. The DMS collects multimodal data across psychological, physical, and physiological measures, supporting both real–time monitoring and feedback during experimentation and reproducible post–hoc analysis and comparative evaluation. Finally, analytical results from the DMS can be fed back into the system to iteratively update scenario and task configurations in the SDC, as well as algorithmic and policy parameters in the EAM, thereby establishing a closed–loop optimization process. Access to the *TransCAVE-E* and related resources is available at https://github.com/TSILAB/TransCAVE-E.

## 3.2 Core Modules

This section further describes the functional composition of the three core modules of *TransCAVE–E* and their respective roles within the overall experimental workflow. *TransCAVE–E* adopts a layered implementation architecture. Unreal Engine and CARLA provide underlying capabilities, including virtual-scene rendering, traffic-agent representation, basic physics simulation, access to simulation states, and network communication. Data acquisition from VR devices, eye-tracking systems, and wearable physiological sensors relies on the corresponding hardware interfaces and software development kits. Building on these fundamental capabilities, the authors developed and integrated the Scenario Design Center, eHMI Algorithm Module, and Data Management System specifically for human–vehicle interaction experiments. The main platform-level contribution of *TransCAVE–E* lies in the unified organization and integration of experimental scenarios and task workflows, external eHMI algorithm input and output, multi-participant experiment control, and multimodal human-factors data management, rather than in reimplementing the underlying rendering engine, general-purpose physics simulation, or device drivers.

### *3.2.1 Scenario Design Center*

The SDC is responsible for translating research questions into executable configurations for virtual experiments. Its primary objective is to support researchers, within a unified platform, in completing traffic scenario construction, task–flow customization, and experimental protocol assignment, thereby providing standardized inputs and an operational foundation for subsequent real–time strategy computation in the EAM and for data acquisition and archiving in the DMS.

Unreal Engine and CARLA provide the underlying virtual environments, traffic agents, physical behaviors, and scene-resource loading capabilities. On this basis, the SDC developed by the authors organizes and encapsulates road environments, traffic facilities, vehicles, pedestrians, and eHMI resources for experimental use. It further provides functions for scenario-parameter configuration, task-flow editing, experimental-condition combination, and experiment-execution management. Its functions can be summarized into four interconnected processes:

(1) Scenario customization: This part integrates environmental models such as road networks and buildings, as well as element models including traffic facilities, vehicle/pedestrian models, and various eHMI assets, enabling researchers to construct high–fidelity virtual traffic environments that are both controllable and reproducible.

(2) Task customization: This part transforms predefined experimental procedures and event scripts into executable task sequences and supports fine–grained configuration of key interaction nodes and conflict situations. In this process, researchers can specify vehicle motion states and behavioral parameters—such as trajectories, speeds, and time headways of controlled vehicles—thereby constructing interaction/conflict scenarios with varying levels of complexity and safety margins.

(3) Experiment customization: SDC also supports combining and packaging edited task sequences under different experimental conditions and managing them as reusable experimental packages.

(4) Experiment execution lobby: Finally, an execution lobby serves as the unified entry point for experiment operation, where researchers can load and launch predefined experimental packages to rapidly deploy experimental workflows under different conditions, and organize participant–specific assignment and implementation.

*3.2.2 EHMI Algorithm Module*

The EAM is the strategy computation and prompt–generation module of *TransCAVE–E*, responsible for transforming real–time interaction states into executable eHMI outputs. Unlike conventional implementations that treat eHMI as a static display variable, the EAM continuously ingests real–time scene data during experiments, dynamically generates prompt commands, and transmits its outputs back to the Server, thereby enabling online validation of eHMI strategies within a closed–loop interaction environment.

Specifically, the Server streams scene and traffic states (e.g., eHMI states, positions of traffic agents, and their relative relations) to the EAM in real time. In AV–pedestrian scenarios, prompts can be triggered based on intent consistency or the degree of conflict, following an adaptive principle of "minimizing disturbance and prompting only when necessary." In HV–AV scenarios, prompts can be further coupled with vehicle interaction strategies to support studies on cooperative driving, conflict avoidance, and trust–related mechanisms. Ultimately, the prompt commands generated by the EAM are written back to the scene by the Server and rendered to pedestrians or drivers in real time via the VR/CAVE Client in visual and/or auditory forms.

*3.2.3 Data Management System*

The DMS constitutes the data component of *TransCAVE–E*'s closed–loop workflow and is responsible for automatic recording and centralized storage of experimental process data. Specifically, during and after experiments, participants' virtual movement trajectories, gaze–point coordinates, SOI fixation statistics, and physiological signals are automatically aggregated into the DMS. Meanwhile, researchers can configure post–trial questionnaires according to the experimental design, which participants complete and submit through the platform, thereby generating subjective scale data. Consequently, the DMS establishes a multimodal data structure across three dimensions, including psychological, physical, and physiological measures, enabling researchers to conduct comprehensive evaluation from both objective behaviors and subjective assessments, while providing a unified data basis for subsequent reproducible analyses and comparative studies.

## 4. Key features

Aiming to support the combined HIL and SIL experiments described above, *TransCAVE–E* incorporates several key features, as illustrated in Fig. 3: (1) highly–customized scenario design, (2) intelligent eHMI incorporation, and (3) distributed multi–agent architecture. This section details these core features.

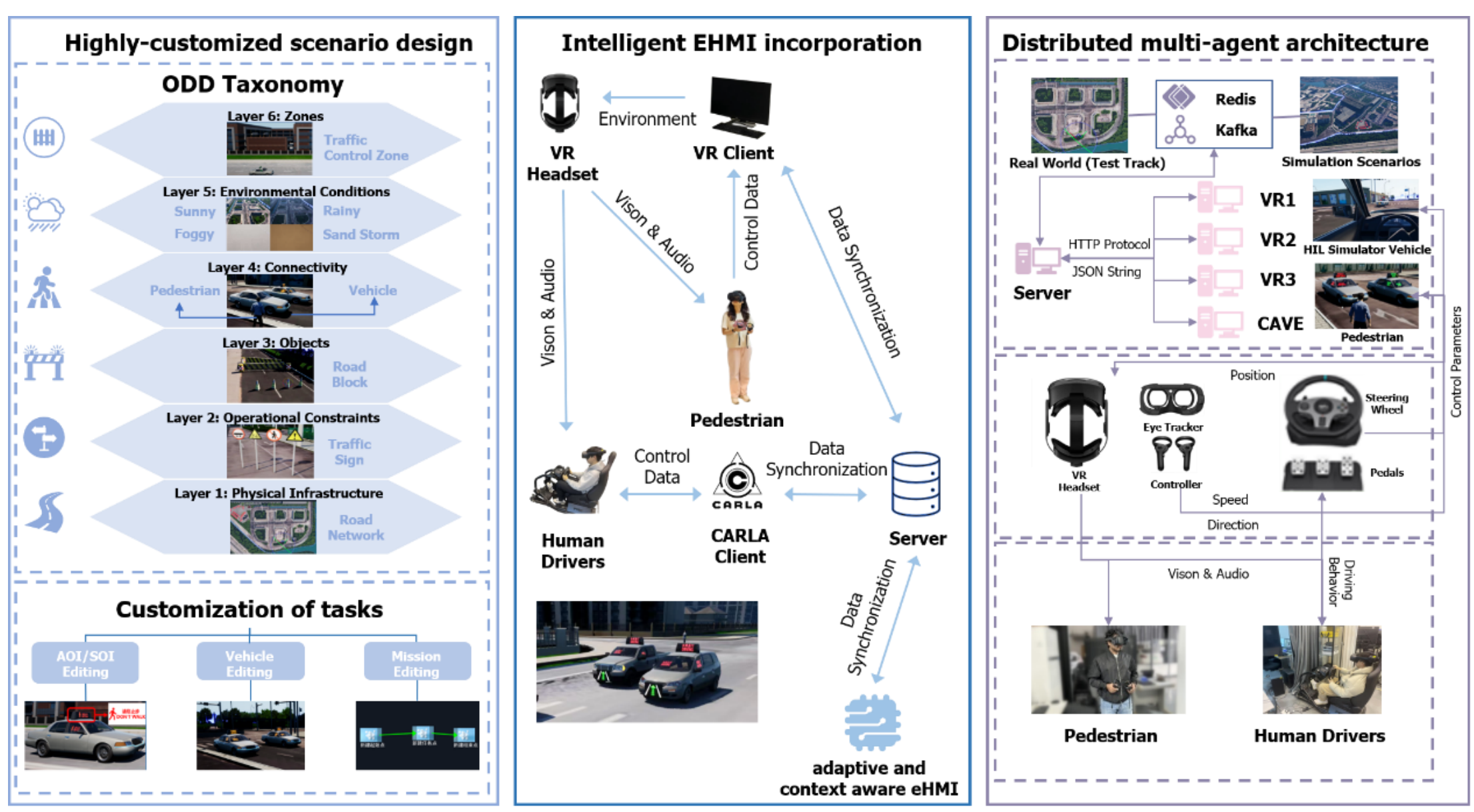

**Fig. 3** Illustration of *TransCAVE–E* features

### 4.1 Highly–customized scenario design

Existing simulation platforms generally provide basic scenario construction capabilities, such as MetaDrive (Li et al., 2022) and Sky–Drive (Huang et al., 2025). However, they remain limited in supporting joint customization across multiple elements and systematic, task–oriented customization (Fremont et al., 2019; Zhang et al., 2020), both of which are particularly valuable for HIL experiments. To address this requirement at the experiment preparation stage, *TransCAVE–E* adopts an integrated design that combines a standard scenario customization system, a task–oriented customization system, and an Area of Interest (AOI)/ Scene of Interest (SOI) editing module, forming a comprehensive toolchain that spans scenario–task customization. The detailed interface contents are shown in Fig. 4. This design enables users to achieve holistic and integrated experiment configurations from both the scenario design and task flow perspectives.

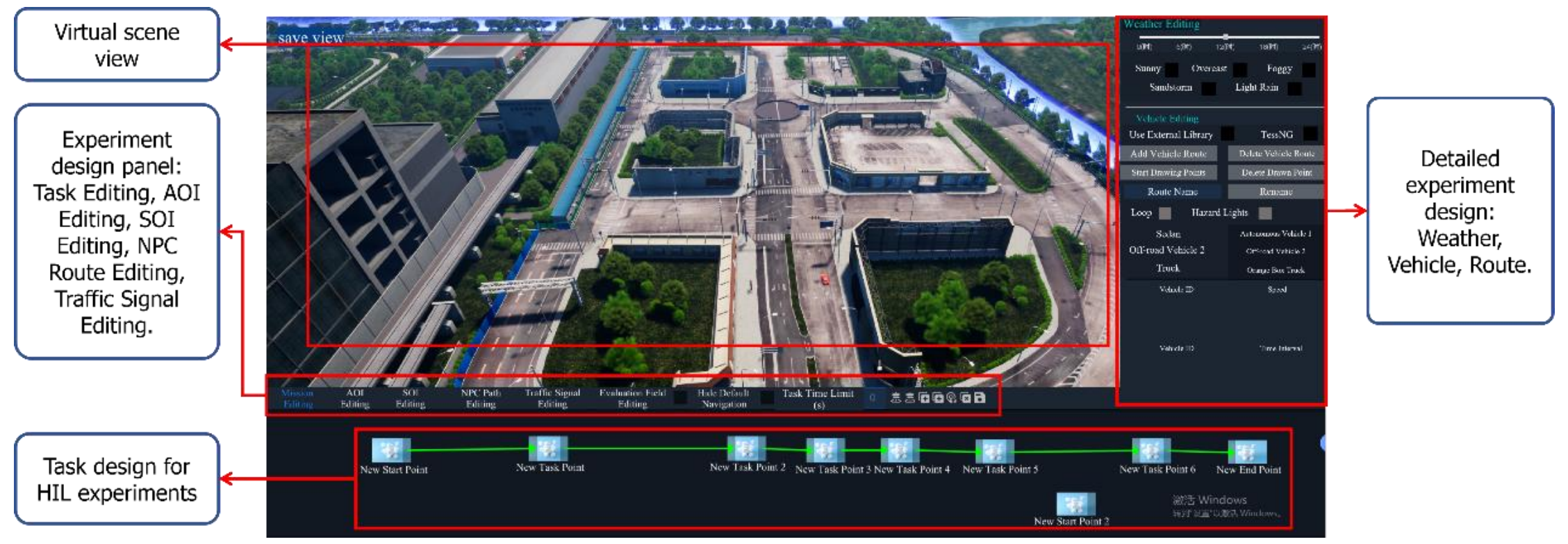

**Fig. 4** Scenario design interface

*4.1.1 Customization of scenarios*

Referring to Operational Design Domain (ODD) design standard (Thorn et al., 2018), the SDC in *TransCAVE–E* is developed based on six modules: Physical Infrastructure, Operational Constraints, Objects, Connectivity, Environmental Conditions, and Zones, enabling in–depth customization of experimental scenarios for complex traffic environments built on Unreal Engine. As illustrated in Fig. 3, users can upload designed road network or 3D scene models to the system, which are then uniformly loaded and managed in SDC. On this basis, the platform allows fine–grained configuration of key environmental factors according to personalized experimental requirements, including Operational Constraints (e.g., traffic signs), Objects (e.g., road blocks), and Environmental Conditions (e.g., sunny, rainy, varying visibility levels). Through these configurations, users can construct complex yet static virtual traffic environments, providing a stable foundation for subsequent human–vehicle and vehicle–vehicle interaction experiments.

*4.1.2 Customization of tasks*

In many HIL studies, volunteers are required to complete a sequence - based experimental procedure. Accordingly, *TransCAVE–E* enables personalized configuration of experimental task workflows and precise deployment of multi–agent interaction logic beyond scenario–level customization. Specifically, the platform supports defining AOI/SOI that trigger predefined incidents, such as appearance of interaction objects, auditory cues or eHMI within the scenario, and organizes these events into executable task sequences. This function allows users to simultaneously control traffic layouts (e.g., vehicle generation and operational logic) and interaction processes (e.g., the temporal coordination between pedestrians/HVs and AVs at conflict points). As a result, *TransCAVE–E* facilitates task - oriented experiment design while maintaining high experimental reproducibility.

### 4.2 Intelligent eHMI incorporation

In addition to the static experiment design, *TransCAVE–E* facilitates dynamic and adaptive eHMI strategies by integrating a unified communication and control framework for both AV–pedestrian as well as HV–AV interactions. As illustrated in Fig. 5, *TransCAVE–E* streams time–synchronized scene and participant states to the EAM, where three complementary capabilities are executed: (1) rule–based prompts; (2) interaction modeling; and (3) context-aware and connected control. The resulting prompt commands and strategy outputs are transmitted back to *TransCAVE–E* and rendered in real time, providing a low–latency testbed for investigating intelligent interaction mechanisms among traffic participants.

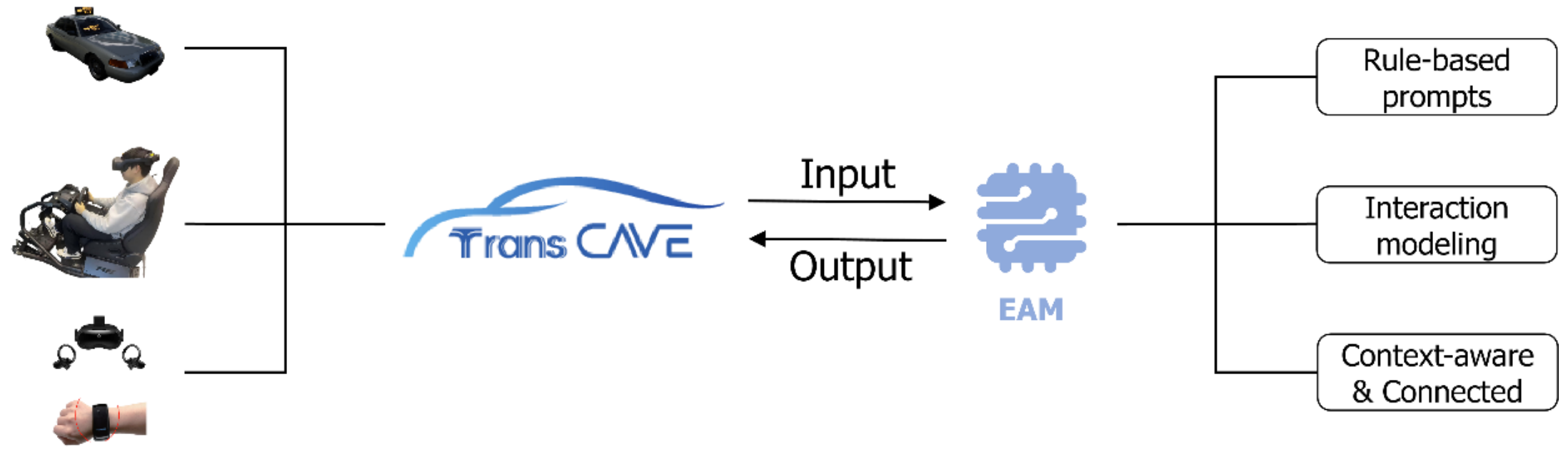

**Fig. 5** Real–time transmission of the eHMI algorithm

#### *4.2.1 Inputs*

*TransCAVE–E* takes virtual environment states and multimodal human–factors data as its primary inputs, enabling simultaneous support for the validation of both AV–pedestrian and HV–AV interaction scenarios.

In AV–pedestrian interaction scenarios, eHMI intent cues are presented through predefined SOI, such as front information panels, light bands, and signal devices mounted on the vehicle's sides or roof. The platform synchronously acquires the eHMI display states, pedestrians' gaze behaviors toward different eHMI regions and physiological signals, as well as pedestrians' movement trajectories in the virtual environment. These synchronized inputs constitute the basis for analyzing how eHMI influences intention comprehension, risk perception, and crossing decision–making.

In HV–AV interaction, human drivers perceive the motion of virtual vehicles and corresponding eHMI cues through immersive VR visualization and generate continuous control inputs via steering wheel and pedals. *TransCAVE–E* records and synchronizes the HV's control behaviors, the dynamically evolving relative distance to the AV, the corresponding eHMI states, and the driver's eye–movement and physiological signals. These inputs support experimental investigations into trust formation, behavioral coordination, and conflict–avoidance mechanisms in cooperative driving.

#### *4.2.2 Outputs*

Based on the above inputs, *TransCAVE–E* outputs two control layers, AV and eHMI control, to enable closed–loop experimentation and evaluation of intelligent eHMI systems. First, the platform outputs AV–level control and state updates within the simulation, including vehicle motion behaviors and scenario–dependent interaction dynamics that participants experience in real time. Second, it outputs eHMI–level control, including eHMI activation and deactivation, display state transitions, and trigger timing, allowing researchers to parametrically manipulate and validate different eHMI designs or strategies under identical scenario conditions. Unlike traditional evaluation paradigms that focus primarily on whether eHMI is noticed or correctly interpreted, *TransCAVE–E* further provides an integrated output representation that links eHMI control states and trigger timing with participants' gaze behaviors, movement trajectories, physiological responses, and post–experiment subjective questionnaire data within a common analytical framework. This unified output supports both quantitative and qualitative comparisons of the cognitive and behavioral effects of different eHMI solutions, thereby enabling systematic optimization, mechanistic interpretation, and validation of intelligent eHMI systems.

### 4.3 Distributed multi–agent architecture

To address the need for real–time interaction among heterogeneous human participants in complex traffic environments (Kalantari et al., 2023), *TransCAVE–E* is designed and implemented with a distributed multi–agent architecture. As illustrated in Fig. 6, this design allows multiple pedestrians and HVs (currently supporting up to four concurrent participants) to operate and interact synchronously within a unified virtual environment, thereby providing essential infrastructure support for multi–user VR experiments.

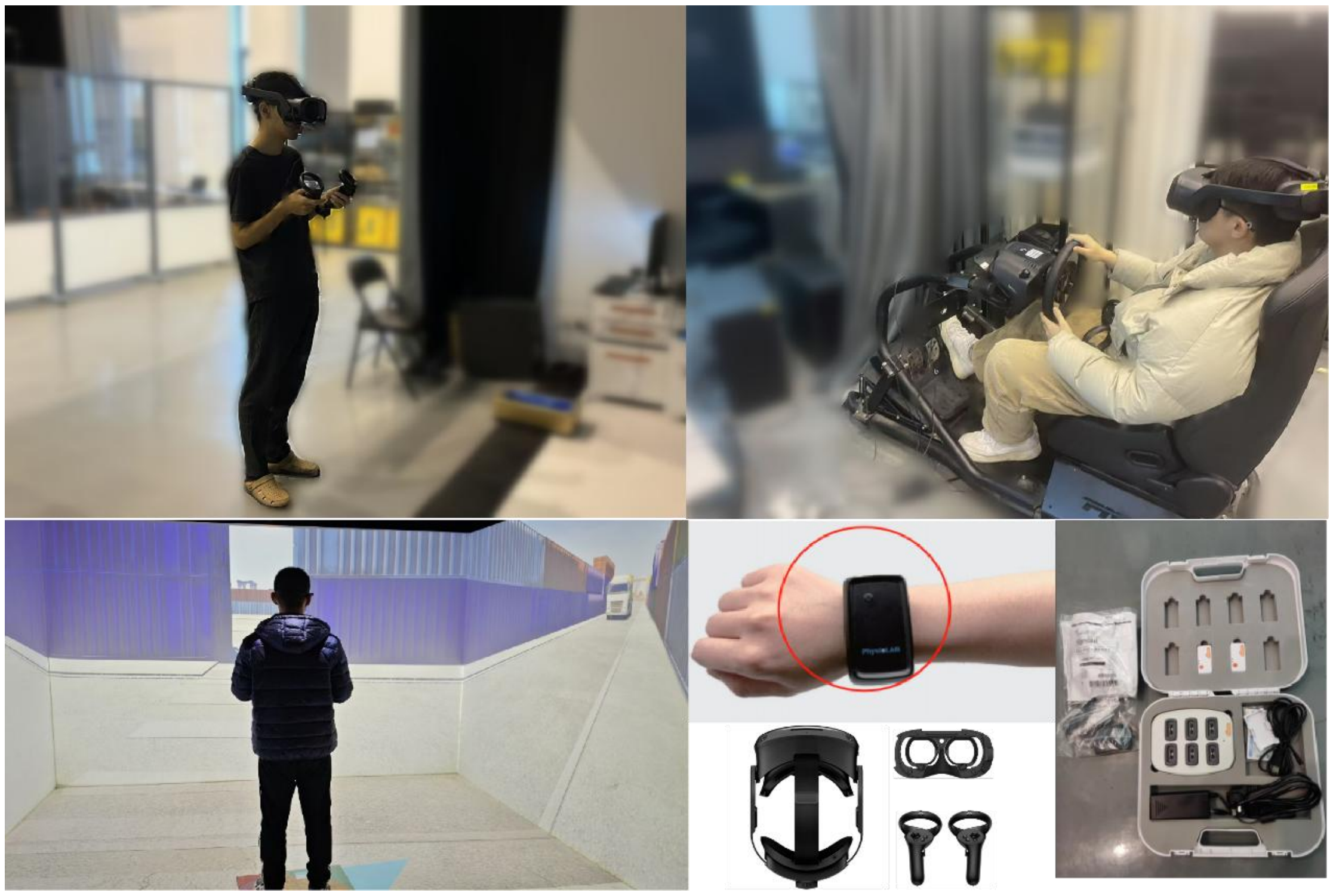

**Fig. 6** Multi–agent experiments: (a) VR headset; (b) Driving simulator; (c) CAVE; (d) Data collection equipment.

*4.3.1 System architecture design*

As illustrated in Fig. 3, the distributed system adopts a Server–Client architecture to jointly support the execution and monitoring of distributed experiments. Specifically, the Server is responsible for unified scheduling and control of the experimental workflow, including scene loading, experiment initiation, and data synchronization. Three VR Clients function as visualization and interaction platforms, which render participants' perspectives in the virtual environment to facilitate online observation and recording of the experimental process. The CAVE client operates as the physiological monitoring Server, responsible for receiving and storing data collected from physiological measurement devices (e.g., VR eye trackers and virtual cockpit systems). Each Client is connected to its corresponding input and sensing devices to control or monitor the state of the associated participant. Meanwhile, all Clients are coordinated by the Server, enabling seamless interaction and state synchronization among different agents.

#### *4.3.2 Communication and sensing techniques*

At the communication level, *TransCAVE–E* employs a dedicated local area network (LAN) to ensure stable connectivity among all devices, thereby maintaining real–time performance and data consistency during multi–agent interactions. The VR scenes developed using UE are rendered through an HTC VIVE Focus 3 headset, and full six–degrees–of–freedom (6–DoF) head tracking is achieved via SteamVR at the Clients, providing an immersive experimental experience.

At the sensing level, the VR headset enables high–precision eye–tracking equipment to capture high–frequency behavioral signals (120 Hz) during experiments, including pupil positions and gaze fixation coordinates. In addition, external wearable sensors, including PhysioWatch and ShimmerStrap, are connected to the Server to synchronously collect multiple physiological signals, such as electrodermal activity, heart rate, and body temperature. These multimodal human – factors data have been proven critical for analyzing participants' attention allocation (Eisma et al., 2019), situational awareness (Kuwata et al., 2026), and variations in cognitive workload (Yang et al., 2025) during complex traffic tasks.

## 5. Use case

So far, *TransCAVE–E* has supported a wide range of use cases that require both HIL and SIL testing. In the HV–AV interaction domain, the platform has been used for closed–loop validation of AV strategies under long–tail scenarios as well as for collecting takeover/failure samples in immersive driving experiments (Liu et al., 2026; Fang et al., 2025). In the pedestrian interaction domain, *TransCAVE–E* enables systematic investigations spanning pedestrian behavioral mechanisms and the effects of eHMI interventions on interaction outcomes (Ye et al., 2026; Ye et al., 2026). Two representative eHMI – incorporated test cases are presented in this section: (i) an intent–recognition–based AV–pedestrian interaction case (Tran et al., 2025), and (ii) a game–theoretic–based HV–AV interaction case (Liu et al., 2026).

### 5.1 AV–pedestrian interaction

The objective of this experiment was to validate the performance of Intent Recognition eHMI (IR–eHMI), an adaptive eHMI system that integrates real–time pedestrian intent recognition and selective eHMI activation strategies (Sun et al., 2025). Fig. 7 illustrates the implementation framework of this case study on the *TransCAVE–E* platform. It should be noted that in more complex multi–vehicle interaction environments, eHMI does not necessarily yield unidirectional benefits and may even induce misleading decisions (Ye et al., 2026). Accordingly, the "selective activation" principle of IR–eHMI constrains prompt triggering to periods characterized by heightened intention conflict and uncertainty, thereby reducing unnecessary informational intrusion and enhancing the contextual adaptability of the strategy.

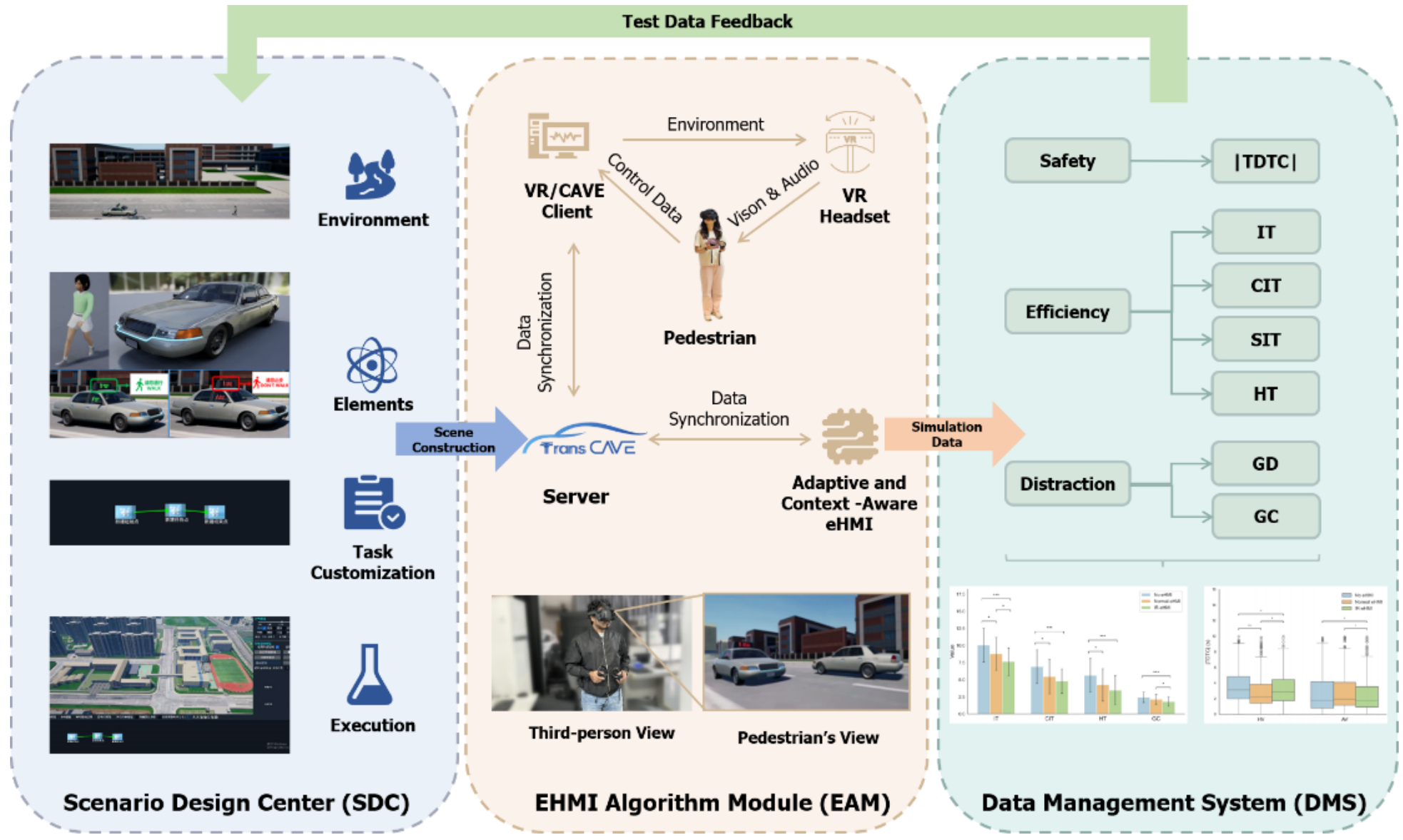

**Fig. 7** Framework to conduct the AV–pedestrian interaction experiment

During the experiment preparation stage, two representative pedestrian crossing scenarios were constructed within the SDC: (i) S1 (AV yielding), in which the AV initially traveled toward the pedestrian from approximately 32 m at 7 m/s, began decelerating at 15 m, and stopped 2.5 m before the conflict point; and (ii) S2 (AV non–yielding), in which the AV started from approximately 35 m away and passed through at a constant speed of 7 m/s. To better reflect real–world traffic complexity, each scenario included an oncoming background vehicle in the opposite lane. Three eHMI conditions were configured, including A (No eHMI); B (Normal–eHMI), in which eHMI was triggered by a fixed distance threshold of 25 m between AV and pedestrian; and C (IR–eHMI), in which eHMI was triggered only when an intention conflict was detected by the IR algorithm and the proposed indicator, CoopScore, fell below a predefined threshold. Based on these configurations, a set of six experimental modules were constructed on *TransCAVE–E* and presented to pedestrian participants in randomized order to support comparative evaluation. The experiment recruited 32 human participants. Each participant completed all six experimental conditions formed by the combination of two vehicle-behavior scenarios and three eHMI modes, and the order of the conditions was randomized for each participant. Before the formal experiment, participants completed eye-tracking calibration and three practice crossings to familiarize themselves with the virtual environment.

The IR algorithm was deployed on the EAM to support real−time eHMI adaptation. In this case, EAM continuously received the speed, location, and distance of the pedestrian and AV from the Server, and dynamically updated the eHMI prompting strategy according to the IR–eHMI logic, which aimed to minimize intrusion and intervene only when necessary. Specifically, CoopScore was updated at a frame–level frequency (10 fps) and used to determine whether eHMI should be activated: when intentions of the AV and the pedestrian were highly aligned (CoopScore $\geqslant$ 0.9), no prompt was issued to avoid unnecessary distraction; when mutual hesitation emerged or both agents tended to proceed first, the CoopScore decreased and corresponding prompts were triggered to improve efficiency or enhance safety. The EAM then continuously transferred the eHMI prompt status (0 for off, 1 for on) back to the Server.

During and after the experiment, the DMS aggregated video recordings, pedestrian and vehicle trajectories, and eye-tracking signals and associated these data with the corresponding scenario, eHMI condition, and experimental trial. The objective measures included the absolute value of Time Difference to Collision (|TDTC|) for safety; Interaction Time (IT), Crossing Initiation Time

(CIT), Stop Initiation Time (SIT), and Hesitation Time (HT) for interaction efficiency; and Gaze Duration (GD) and Gaze Count (GC) for visual distraction. Post-trial questionnaires further assessed perceived safety, the timing of intention recognition, attention to the surrounding traffic environment, and perceived eHMI effectiveness.

As reported in the original study, overall differences among the three eHMI conditions were examined using one-way analysis of variance (ANOVA), followed by pairwise independent-samples t-tests to identify specific between-condition differences, with statistical significance defined at $p < 0.05$. In this case, the DMS supported the joint organization of behavioral measures, eye-tracking indicators, and subjective assessments within the same experimental-condition structure.

Based on data from 32 participants, the reported analyses showed that IR–eHMI improved pedestrian decision efficiency by 12.8% in S1 and 13.0% in S2, reduced gaze distraction by 17.1% in S1, and reduced the number of eHMI prompts by 40% in S2 while maintaining overall interaction safety. From a platform perspective, this case demonstrates a continuous experimental workflow in which the SDC combines vehicle behaviors and eHMI conditions into reproducible task modules, the EAM executes an external intent-recognition algorithm and returns eHMI activation commands to the virtual environment, and the DMS links the resulting eHMI states with trajectories, gaze measures, and subjective assessments.

### 5.2 HV–AV interaction

Driver interaction testing constitutes a critical pathway for the evaluation and data acquisition of long–tail, safety–critical scenarios. To further evaluate *TransCAVE–E*'s capability to support HV–AV simulation scenarios, Liu et al. (Liu et al., 2026) conducted an additional use case on validating eHMI information disclosure strategies in HV–AV unprotected left–turn conflicts. This case focuses on a particularly challenging HV–AV interaction context—the unprotected left turn. Fig. 8 illustrates the implementation framework of this case study on the *TransCAVE–E* platform.

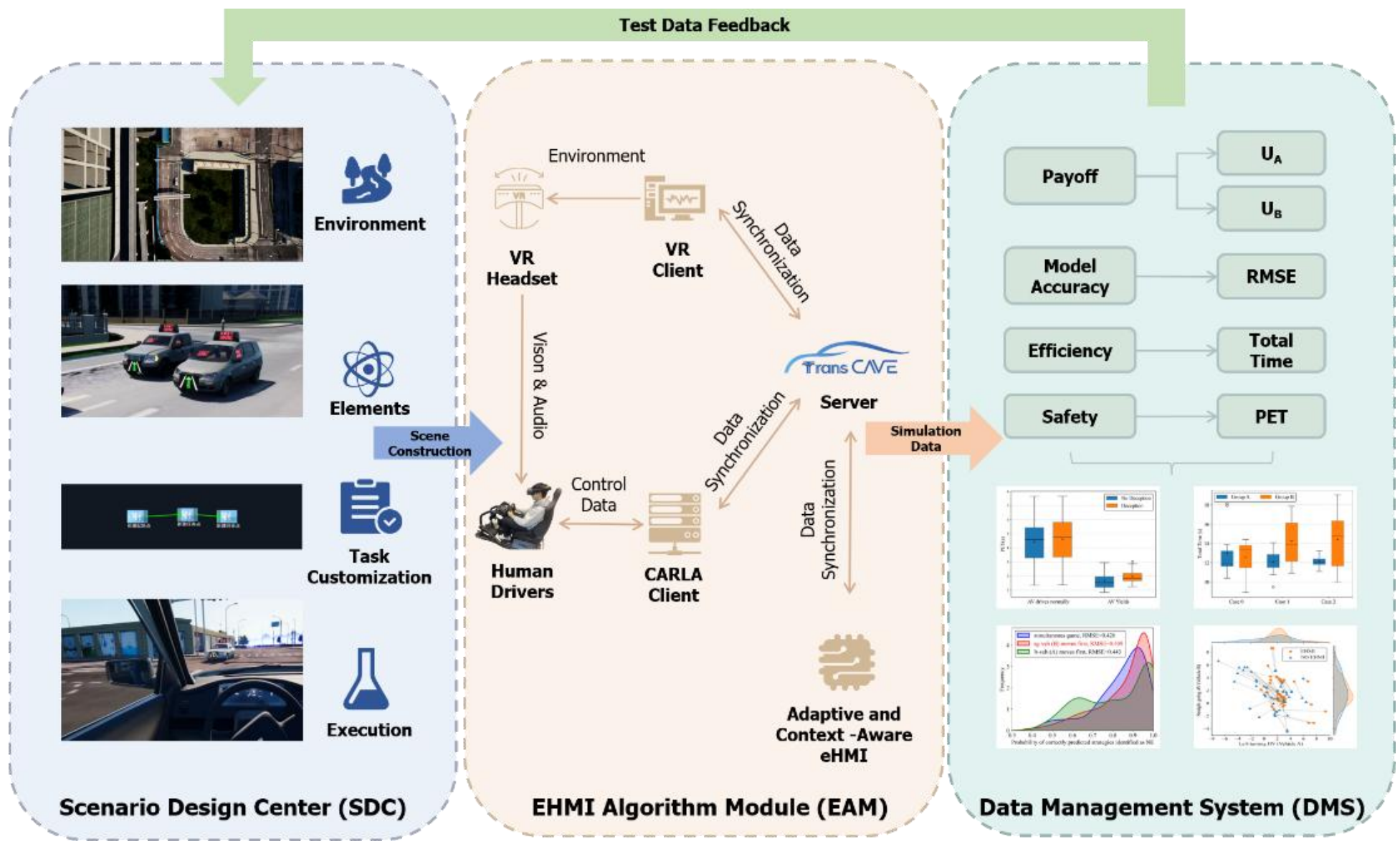

**Fig. 8** Framework to conduct the HV–AV interaction experiment

Within the SDC, we instantiated an unprotected left–turn conflict scenario in which the left–turning vehicle is a HV, while the through–moving vehicle is an algorithm–controlled AV equipped with eHMI; participants executed the left–turn maneuver along a predefined route and traversed the conflict point. Three experimental conditions were configured: Case 0, no eHMI; Case 1, eHMI discloses fully truthful information; and Case 2, eHMI follows an information disclosure framework that allows non–fully–truthful disclosure. In addition, we introduced a trust–context manipulation in the procedure: Group A was not informed prior to the experiment that deceptive disclosure might occur (representing a higher–trust baseline), whereas Group B was informed in advance that the eHMI might provide false information (representing a reduced–trust condition). The platform further constrained the simulation and prompt refresh rates to balance perceptibility with cognitive workload.

Within EAM, the HV–AV interaction was formalized as a signaling game with an explicit information disclosure stage. Unlike conventional formulations that primarily characterize the sequential actions of the interacting agents, the signaling–game formulation explicitly introduces a disclosure stage and allows both truthful and deceptive disclosures to coexist. The algorithm decomposes the disclosure decision into three subproblems, whether to disclose, when to disclose, and which intention information to disclose, and optimizes toward an "expected strategy" that maximizes expected payoff and safety. Through eHMI, the AV influences the HV's subsequent strategy choice: under the assumption that the HV tends to trust the disclosed information, the HV selects the action that maximizes its perceived utility. This mechanism yields the possibility of prosocial deception under certain conditions, where the disclosed intent is inconsistent with the AV's eventual execution but is intended to induce a safer interaction outcome.

After the experiment, the DMS aggregated multimodal records, including driving videos, vehicle trajectories, and system logs of the eHMI disclosure states under different experimental conditions (Case 0–2) and trust contexts (Group A/B). Interaction performance was quantified for both vehicles using payoff functions ( $U_A$ , $U_B$ ), enabling payoff–based comparisons across different disclosure strategies. Model accuracy was evaluated using the root mean square error (RMSE) between observed strategy choices and model predictions during the calibration or validation stage, providing an error–based check of how accurately the game–theoretic model reproduces experimental behavior. Efficiency was measured by the total time required for both vehicles to clear the conflict point, while safety was assessed using PET. In addition, a post–condition questionnaire captured subjective trust in the eHMI (rated on a 0–10 scale), providing complementary evidence to characterize how disclosure strategies relate to trust and subsequent driving behavior.

The study suggests that, in some hazardous interactions, the game–theoretic disclosure framework admits both theoretical and experimental support for the potential safety benefits of prosocial deception; moreover, when human drivers' trust decreases, eHMI usage may negatively affect interaction efficiency, indicating that the coupling between disclosure policy, human trust, and efficiency consequences should be treated as a key design constraint. Importantly, the implementation of this use case demonstrates that *TransCAVE–E* can support the execution of driver interaction experiments and SIL coupled validation of eHMI strategies (including game–driven information disclosure policies), thereby providing an extensible and reproducible integrated experimental infrastructure for the design and evaluation of intelligent eHMI in HV–AV interaction scenarios.

## 6. Conclusion

This study presented *TransCAVE–E*, a distributed simulation-based experimental platform developed to support the transition of eHMI research from predefined interface evaluation toward closed-loop testing of adaptive and intelligent communication strategies. The platform addresses the gap between study-specific human-factors experimentation and algorithm-oriented automated-driving simulation by integrating experiment orchestration, human-in-the-loop interaction, software-in-the-loop strategy execution, and multimodal human-factors evaluation within a common experimental infrastructure.

*TransCAVE–E* organizes this workflow through three tightly coupled modules and a distributed interaction architecture. The SDC translates research questions into reusable scenario, task-flow, and experimental-condition configurations; the EAM establishes bidirectional real-time communication between the simulation and independent external eHMI algorithms; and the DMS synchronizes and associates behavioral, eye-tracking, physiological, system-log, and subjective data with the corresponding experimental conditions and trials. The distributed multi-agent architecture further enables heterogeneous human participants and traffic entities to interact synchronously in a shared virtual environment. The principal contribution of the platform therefore lies not in replacing existing rendering engines, physics simulators, or sensing devices, but in integrating these underlying capabilities into an experiment-oriented HIL–SIL workflow for intelligent eHMI research.

Two use cases demonstrated the applicability of this workflow to different interaction mechanisms and participant roles. In the AV–pedestrian case, an external intent-recognition algorithm was executed through the EAM to dynamically control selective eHMI activation. With 32 participants, the IR–eHMI improved decision efficiency by 12.8% and 13.0% across the yielding and non-yielding scenarios, respectively, reduced gaze distraction by 17.1% in the yielding scenario, and reduced the number of eHMI prompts by 40% in the non-yielding scenario while maintaining overall interaction safety. More importantly from a platform perspective, the case demonstrated continuous coupling among reusable task configuration, external-algorithm execution, dynamic eHMI control, and multimodal behavioral evaluation. The HV–AV case further extended the workflow to active driver interaction by integrating a signaling-game-based information-disclosure strategy in an unprotected left-turn conflict. The platform supported real-time strategy execution together with vehicle-state recording, trust manipulation, safety and efficiency evaluation, illustrating its applicability to strategy-level SIL testing beyond pedestrian-oriented eHMI experiments.

Several limitations remain. The current distributed implementation supports up to four concurrent human participants, and communication and synchronization overhead may become a bottleneck as the number of users and connected devices increases. Future work should therefore investigate more scalable communication and synchronization architectures for larger multi-agent experiments. In addition, although immersive VR provides strong experimental controllability and enables the systematic study of safety-critical interactions, differences remain between virtual experiments and real-world traffic in environmental fidelity and behavioral validity. Future development should consequently combine controlled VR experimentation with full-scale or real-world validation to quantify cross-environment consistency and improve the external validity of platform-based findings. Addressing these challenges will further strengthen *TransCAVE–E* as an extensible infrastructure for the development, validation, and iterative refinement of intelligent human–vehicle communication strategies.

**Authorship contribution statement**
**Yun Ye:** Conceptualization, Formal analysis, Project administration, Investigation, Methodology, Validation, Funding acquisition, Writing-original draft. **Zexuan Li:** Methodology, Data curation, Formal analysis, Investigation, Visualization, Writing-original draft. **Haoyang Liang:** Conceptualization, Methodology, Resources, Writing-review & editing, Funding acquisition. **Boya Sun:** Data curation, Formal analysis, Writing-review & editing. **Jian Sun:** Conceptualization, Supervision, Resources, Funding acquisition, Writing-review

& editing. **Haotian Shi:** Conceptualization, Validation, Resources, Writing-review & editing.

## Declaration of the use of AI

During the preparation of this work the authors used ChatGPT in order to improve the readability and language of the manuscript. After using this tool/service, the authors reviewed and edited the content as needed and take full responsibility for the content of the publication.

## Ethical statement

All human-involved experiments included in this study were approved by the Ethics Committee at Tongji University (Approval No. tjdxsr2024041 and No. tjdxsr2025011). Informed consent was obtained from all individual participants included in the study.

## Data availability statement

Data will be made available on reasonable request. The accessible URL is https://osf.io/g2mv4/overview.

## Declaration of competing interest

The authors declare that they have no known competing financial interests or personal relationships that could have appeared to influence the work reported in this paper.